\documentclass[onecolumn]{aa_mod}

\usepackage{graphicx}
\usepackage{txfonts}

\usepackage{rotating}                  

\begin{document}
   \title{Astrocladistics: a phylogenetic analysis of galaxy evolution}

                      \titlerunning{Astrocladistics: a phylogenetic analysis of galaxy evolution II} 
                      \authorrunning{Fraix-Burnet et al.} 
                      
   \subtitle{II. Formation and diversification of galaxies}

   \author{Didier Fraix-Burnet \\ \textit{Laboratoire d'Astrophysique de Grenoble, France }
          \and \\ \vskip 0.1 true cm
          Emmanuel J.P. Douzery \\ \textit{Laboratoire de Pal\'eontologie, Phylog\'enie et Pal\'eobiologie, 
          \\ Institut des Sciences de l'\'Evolution de Montpellier, France}
          \and \\ \vskip 0.1 true cm
          Philippe Choler \\ \textit{Laboratoire d'\'Ecologie Alpine, Grenoble, France}
          \and \\ \vskip 0.1 true cm
          Anne Verhamme \\ \textit{Laboratoire d'Astrophysique de Grenoble, France}
          }
\institute{
             \footnotetext{Anne Verhamme is now at Geneva Observatory, Switzerland. First author's address: Laboratoire d'Astrophysique de Grenoble, BP 53, F-38041 Grenoble cedex 9, France, email: fraix@obs.ujf-grenoble.fr. Supplementary material (numerical tables and character projections) is available on http://hal.ccsd.cnrs.fr/aut/fraix-burnet or http://www-laog.obs.ujf-grenoble.fr/public/fraix.}
 }            

\date{}

   \abstract{ This series of papers is intended to evaluate astrocladistics in reconstructing phylogenies of galaxies. The objective of this second paper is to formalize the concept of galaxy formation and to identify the processes of diversification. We show that galaxy diversity can be expected to organize itself in a hierarchy. In order to better understand the role of mergers, we have selected a sample of 43 galaxies from the GALICS database built from simulations with a hybrid model for galaxy formation studies. These simulated galaxies, described by 119 characters and considered as representing still undefined classes, have experienced different numbers of merger events during evolution. Our cladistic analysis yields a robust tree that proves the existence of a hierarchy. Mergers, like interactions (not taken into account in the GALICS simulations), are probably a strong driver for galaxy diversification. Our result shows that mergers participate in a branching type of evolution, but do not seem to play the role of an evolutionary clock. 

    \keywords{Cladistics --
              Galaxies: fundamental parameters --
              Galaxies: evolution --
              Galaxies: formation --
              Methods: numerical
              }
   }

   \maketitle
   \newpage
%

\section{Introduction}

When Hubble discovered the true nature of galaxies in 1922 (Hubble~\cite{h22}), galaxy diversity was limited to morphological differences and four types were enough to describe these new objects. Nowadays, galaxies are known to be complex and diversity is exposed through characteristics of numerous observable parameters. Physics and chemistry of the basic constituents (stars, gas and dust, see \cite{paperI}) and their interactions have been observed and modelled (Vilchez, Stasinska and Perez~\cite{evol1}, Sauvage, Stasinska and Schaerer~\cite{evol2}, Hensler, Stasinska, Harfst, Kroupa, and Theis~\cite{evol3}). The traditional, ever revised, Hubble classification (Hubble~\cite{h26}, Hubble~\cite{h36}, de Vaucouleurs~\cite{devauc}, Sandage~\cite{sandage}, Roberts \& Haynes~\cite{rh}, van den Bergh~\cite{vdb}) is no longer suited to encompass all this variety, notably at high redshifts where galaxies do not look like those in our neighbourhood (e.g. van den Bergh~\cite{vdb}).
In addition, numerical simulations somewhat enlighten the interplay between all constituents of galaxies and some of their behaviour with time (Hatton, Devriendt, Ninin, Bouchet and Guiderdoni~\cite{galics1}, Bournaud , Combes and Jog~\cite{bournaud}, Menci, Cavaliere, Fontana, Giallongo, Poli and Vittorini~\cite{menci}). However, because of galaxy complexity, they are necessarily incomplete, in particular concerning morphological aspects, so that the inadequate classification makes a comparison between simulated and true samples difficult. 

In 1936, Hubble imagined that galaxies should evolve (Hubble~\cite{h36}). In this way, he explained the origin of spiral galaxies as being relaxed ellipticals. While galaxy evolution is now universally recognized, historical scenarios as well as formation and nature of the very first objects are far from being established (e.g. Bromm, Ferrara, Coppi and Larson~\cite{firststars}, Bromm and Loeb~\cite{firstlight}, Schneider, Ferrara, Natarajan and Omukai~\cite{firstsobj}). It is often understood that the formation of a galaxy deals only with its first appearing in the Universe as an entity and that afterwards it evolves with more major or less major modifications. We believe that this definition of formation is not adapted to a diversity generated in the course of evolution. This concept should be formalized more clearly when dealing with phylogeny (``species evolution'') of galaxies (see Sect.~\ref{formconcept}).

To address these two questions (classification and formation -- evolution), Fraix-Burnet, Choler and Douzery  (\cite{FCDa}) proposed to use cladistics, a methodology borrowed from evolutionary biology (see also Fraix-Burnet~\cite{fraix}). In the present two companion papers, we detail the fundamentals of astrocladistics. The first paper (\cite{paperI}) concentrates on the applicability of cladistics to objects like galaxies, emphasizing also the practical course of the analysis. Readers are referred to this paper for principles and more details of cladistics. The present second paper is devoted to a conceptualization of the formation and evolution of galaxies as a whole, that is required to understand the organization of their diversity. We also analyse a sample of simulated galaxies in order to study the specific role of mergers.

In this paper as in \cite{paperI}, we use the word ``class'' in a generic manner, and avoid the word ``type'' because it is inevitably linked to a Hubble morphological classification. We also make the one object -- one class assumption which means that an individual galaxy of the simulation is supposed to be generic for a class. This has the advantage of more generality because even though with our sample of simulated galaxies we are able to identify the fate of a galaxy at different epochs, this is not possible with real objects. Like in \cite{paperI}, we do not yet specify any distinction between ``class'' and ``species'' for galaxies.

In Sect.~\ref{formdiv}, we present some general concepts of evolution, and detail the formation and diversification processes for galaxies. The selection of the sample of simulated galaxies is described in Sect.~\ref{galmerg} and the analysis inputs are listed in Sect.~\ref{cladanal}. Results are presented in Sect.~\ref{results}, while a discussion can be found in Sect.~\ref{discussion}. The conclusions are given in Sect.~\ref{conclusions}.


\section{Formation, diversification and hierarchy}
\label{formdiv}

\subsection{Descent with modification}
\label{biology}

The hierarchical organization of the diversity of living organisms is an empirical result found already at the time of Aristotle, and thoroughly established in the 18th century when the - now universal - Linn\'e nomenclature came into use (Knapp~\cite{knapp}, \cite{paperI}). Darwin (\cite{darwin}), in his theory of evolution, explained this tree-like structure by a descent with modification process: each organism transmits its characteristics to a descendant with or without modification. Descent explains pattern similarity while modification explains pattern differences. Because these modifications can be numerous for such complex systems, a given species generally evolves into several new species. This is called branching evolution and cladistics is designed to find such hierarchies (Hennig~\cite{hennig}, Wiley, Siegel-Causey, Brooks and Funk~\cite{cladist}, see Paper~I).

In the biological world, another component of evolution tends to alter the tree-like structure. In some cases, genes are directly transmitted from one organism to another, creating a hybrid organism with characteristic patterns of both original species. This is called reticulated evolution because on a tree it would appear as a junction between two different branches (e.g. Woese~\cite{woese}, Legendre~\cite{legendre}). When branching evolution is dominant, reticulated evolution brings some noise in the cladistic analysis and hybridization events can be identified. The situation is more difficult when the reverse is true, as seems to be the case for bacterial evolution (Doolittle~\cite{doolittle}).

It is important to point out that classification here does not concern individual objects but species (Wiley et al.~\cite{cladist}). A chimpanzee does not evolve into a human, but both species have a common ancestral species (often simply called ancestor). In biology, a given species slowly evolves through many generations of individuals. Differences are tiny at each generation, but after a while new individuals depart significantly enough for a new species to be defined. The notion of species is not unique and is not defined only by interbreeding (see Brower~\cite{brower2000} and some references therein). For instance, it can be defined from cladograms, that is from trees produced by a cladistic analysis, in which case it is coined a 'clade'. As explained in \cite{paperI}, objects thus grouped share a common ancestor which has transmitted its evolved ('derived') characteristics (innovations of evolution).

In astrophysics, as will be seen in Sect.~\ref{diversification}, the evolutionary process is much more spectacular. The term 'progenitor galaxies', often used, means in reality 'progenitor class to a given class of galaxies'. This point, very important for astrocladistics, is also discussed and illustrated in \cite{paperI}, and formalized further in Sect.\ref{formconcept}. 

\subsection{Formation of galaxies}
\label{formconcept}

   \begin{figure}
   \centering
   \includegraphics[width=15 true cm]{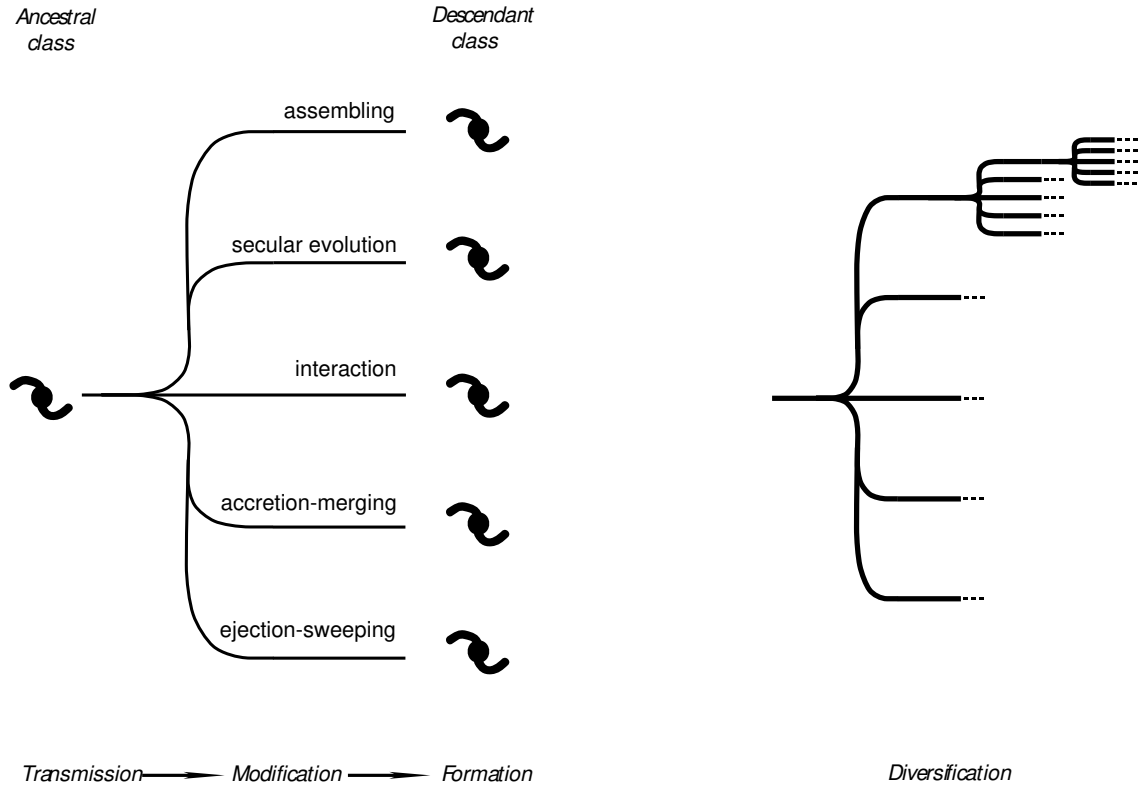}
      \caption{Formation processes of galaxies (left) and diversification scheme (right).
              }
         \label{formation}
   \end{figure}

It is often implicitly admitted that a galaxy forms at an early epoch of the Universe and evolves to the present time (e.g. Steidel~\cite{steidel}). In other words, ``formation'' and ``evolution'' are disconnected. However, we are interested in understanding how galaxies happen to be \emph{as we see them}, that is how they formed themselves into the way they are in our data. As already mentioned, we are concerned with the class a given galaxy represents rather than with the individual by itself. Hence, any process affecting some properties of a galaxy is a formation process. In this sense, formation is to be understood as formation of a given class of galaxies. Five processes that change galaxy characteristics can be identified:

\begin{enumerate}
	\item Even though the nature of the very first objects is still under debate (e.g. Bromm et al.~\cite{firststars}, Bromm \& Loeb~\cite{firstlight}, Schneider et al.~\cite{firstsobj}), some basic constituents (stars, gas and dust, see \cite{paperI}) assembled themselves in a self-gravitational entity called a galaxy. This simple scheme renders possible the appearing of a new galaxy at any epoch of the Universe. 
	\item Galaxies live in an environment made up of gas and a gravitational field shaped by dark matter and other galaxies. However, for a given period, a galaxy can be isolated or located in a stable environment. Yet, as shown in Paper~I, its basic constituents (mainly stars, gas and dust) evolve, hence the galaxy evolves. For instance, if a sufficient number of supernovae explode, the global metallicity of the intragalactic gas is increased, so that a new class of galaxies could be defined. Even though this secular evolution is less spectacular than the four other ones, it is undoubtedly a formation process as well. In \cite{paperI}, it is shown that secular evolution alone yields diversity organized in a hierarchy.
	\item Interactions between galaxies (or between a galaxy and any gravitational perturbation) are probably very frequent, and generally very violent (e.g. Menci et al.~\cite{menci}). Internal kinematics is necessarily affected, triggering starbursts, chemical reactions in the gas and dust, structural and morphological changes, nuclear activity, feeding of black holes, and so on. After an interaction, a galaxy is certainly different, belonging to a new class, so that interactions are a formation process and a strong driver of diversity.
	\item Sometimes, the consequence of an interaction is merging. For instance, in the case of the merging of two spiral galaxies with comparable masses giving birth to an elliptical galaxy, two galaxies obviously disappear and a new one, very different, is born. Such major merger events clearly constitute a formation process. If the masses are very different, we could consider that only the small galaxy has disappeared, eaten or accreted by the bigger one. But still, the latter is modified (at least its mass has increased, but other characteristics are probably modified as well) and could lead to the definition of a new class of galaxies. These minor mergers can be viewed as accretion, like that of intracluster gas or whatever. We consider the accretion-merging formation process globally and discuss it in more detail in Sect.~\ref{brevme}.
	\item Ejection or sweeping of material from a galaxy is just the opposite phenomenon to accretion-merging, less violent in general however, and is similarly a formation process.
\end{enumerate}

These five formation processes: assembling, secular evolution, interaction, accretion-merging, ejection-sweeping are depicted in Fig.~\ref{formation}. This makes it clear that galaxies are always formed from material coming from other galaxies or from the intergalactic medium. In all cases, the building blocks have their own history, they have evolved themselves. Formation necessarily includes evolution, even in the extragalactic world. Stellar astrophysicists are used to this kind of transmission to a new generation or to a new class of objects. There are several populations of stars, the more recent (containing more metals) being made up of material processed in stars of the previous population and ejected in the interstellar medium. In the case of galaxies, this mechanism is reminiscent of the ``descent with modification'' of the Darwinian evolution: basic constituents of galaxies are transmitted, and modified through the five formation processes, to a newly formed descendant.

\subsection{Diversification of galaxies}
\label{diversification}

After a galaxy of a given class appears, it is affected by one of the processes stipulated above, then yields a descendant that belongs to a new or the same class. This is very different from species evolution in biology as already said in Sect.~\ref{biology} because this happens only very gradually after a large number of generations. However, very rarely, some individual living organisms have to give birth to descendants of a different species. For galaxies, an individual can be so much transformed by processes identified in Sect.~\ref{formconcept} that a new galaxy belonging to a different class is formed. The diversification cycle is then at work (see Fig.~\ref{formation}). The goal of a phylogenetic analysis of galaxy evolution like astrocladistics is to classify galaxies according to their history of formation processes. An observed galaxy is the result of several formation events, and the analysis should later be able to establish whether the order in which of they have occurred matters or not.

It is important not to be confused between number of classes (diversification) and number of galaxies. Even if gravity tends to diminish the total number of galaxies through mergers, diversification (i.e. number of different classes of objects) increases because galaxies are complex objects: four perfectly identical galaxies merging by pairs will yield two objects certainly different from each other and from the original ones. Even if at the very last we end up with one object, it will characterize a new class, making altogether four classes in this simple example. Hence diversification still occurs even if the Universe should finish in a Big Crunch (but this does not seem to be the case: Benoit et al.~\cite{archeops}, Spergel et al.~\cite{wmap}).

One necessary condition for diversification to occur in a ``descent with modification'' scheme is that the number of generated species be sufficient. This is probably not the case for stars. But among the galaxy formation processes identified above, assembling is the rarest of all, appearing only once for any full lineage of galaxies. Its time scale can thus be set at the Hubble time ($t_H$) which is the age of the Universe currently estimated at $13.7\ 10^9$~years (e.g. Spergel et al.~\cite{wmap}). Modifications through secular evolution are permanent, but those implying class changes can probably be estimated to occur at the rate $t_H / n_{se}$ where $n_{se}\simeq$ 2 to 3. This means that if a galaxy and its descendants remain isolated during the age of the Universe, they could globally belong to 2 or 3 different classes. Interactions are clearly the most frequent events in the lives of galaxies being always plunged into the gravitational field which is shaped by other galaxies and dark matter. A quite conservative value for its occurrence rate would be something like $t_H / n_{in}$ with $n_{in}\simeq$ 10 to 50. Accretion-merging is slightly less frequent, and from the GALICS simulations (see Sect.~\ref{galmerg}), we estimate the rate at 
$t_H / n_{ac}$ with $n_{ac}\simeq$ 5 to 10. Ejection-sweeping is a rather rare event, that can be attributed the same rate as secular evolution: $t_H / n_{ej}$ where $n_{ej}\simeq$ 2 to 3. 

Hence, even if the above guesses are very approximative, the number of transformations caused by all the formation processes is certainly sufficient for quite a few classes to appear. Astrocladistic analyses will reveal whether they are organized in a hierarchy or not.

We must stress that the hierarchical organization of galaxy diversity considered in astrocladistics has nothing to do with the hierarchical scenario of galaxy formation that deals solely with their mass distribution (e.g. Hatton et al.~\cite{galics1}) and is depicted by an inverse tree (a lot of small galaxies merging into a fewer big ones).

\subsection{Branching evolution and mergers}
\label{brevme}

Except for merging, all the processes described above (Fig.~\ref{formation}) obviously produce branching evolution, leading to a hierarchical organization of the diversity: a given class gives birth to at least a new one. 
When two galaxies of two different classes merge, a kind of hybrid object is produced by mixing together their basic constituents. This could a priori parallel reticulated evolution (Sect.~\ref{biology}). 
It is thus possible that the historical information could be lost or impossible to extract from observations. 
However, since interactions are probably the dominant process of galaxy diversification, mergers are expected to bring only some noise in the cladistic analysis. In addition, major mergers (mass ratio lower than 1/3) are rarer than minor ones which somehow can be compared to accretion. Finally, major mergers are remarkable because they yield an elliptical galaxy from two spirals in numerical simulations (e.g. Barnes~\cite{barnes}, Bournaud et al.~\cite{bournaud}). This is based on morphological considerations only. In reality, the velocity dispersion is dramatically increased, forming a bulge from the disks, and starbursts certainly appear as well. Considering all properties of galaxies, these events might not look so catastrophic compared to other events, particularly to interactions.

Moreover, galaxies are not living organisms and their constituents do not behave like genes. In biological hybridization, a gene of one species replaces a gene of the other species. For galaxies, it is always a mixture of basic constituents. 

The conclusion is that mergers do not seem to be able to destroy a hierarchical organization of galaxy diversity, and might quite possibly participate in it. We propose in the next sections to test this point by performing an astrocladistic analysis on a sample of simulated galaxies with different numbers of merger events in their entire history. If a tree is found, then it will be possible to check whether the number of mergers could be a kind of cosmological clock for galaxy evolution, whether mergers are the main diversification driver among the formation processes taken into account in the simulations (i.e. assembling, secular evolution and accretion-merging).

\section{Selecting the sample}
\label{galmerg}

GALICS (Galaxies In Cosmological Simulations) is a hybrid model for hierarchical
galaxy formation studies, combining the outputs of large cosmological N-body simulations
with simple, semi-analytic recipes to describe the fate of the baryons within dark matter haloes (Hatton et al.~\cite{galics1}, \cite{paperI}). As hot gas cools and falls to the centre of these haloes, it settles
in a rotationally supported disc. Galaxies remain disc structures unless mergers or instabilities occur, in which case simple recipes are used to develop a bulge and a burst components. Hence a galaxy is described by these three components each one having its own parameters of geometry, dynamics, masses (stars and gas), metallicity and photometry from the ultraviolet to the far infrared.
Assembling (first appearing in the simulation as a minimum local overdensity), secular evolution and accretion-merging are taken into account. Interactions between galaxies and ejection-sweeping phenomena are not considered. 

\begin{table}
	\begin{center}
		\begin{tabular}{rlrlrl}

  1& bol\_lum                 &  2& IR\_bol                   &                               \\
   &                          &   &                           &                               \\
  3& disc\_mgal               & 11& bulge\_mgal               & 19& burst\_mgal               \\
  4& disc\_mcold              & 12& bulge\_mcold              & 20& burst\_mcold              \\
  5& disc\_mstar              & 13& bulge\_mstar              & 21& burst\_mstar              \\
  6& disc\_mcoldz             & 14& bulge\_mcoldz             & 22& burst\_mcoldz             \\
  7& disc\_rgal               & 15& bulge\_rgal/disc\_rgal    & 23& burst\_rgal               \\
  8& disc\_speed              & 16& bulge\_speed              & 24& burst\_speed              \\
  9& disc\_tdyn               & 17& bulge\_tdyn               & 25& burst\_tdyn               \\
 10& disc\_inst\_sfr          & 18& bulge\_inst\_sfr          & 26& burst\_inst\_sfr          \\
   &                          &   &                           &   &                           \\
 27& disc\_JOHNSON\_U*        & 58& bulge\_JOHNSON\_U*        & 89& burst\_JOHNSON\_U*        \\
 28& disc\_JOHNSON\_B*        & 59& bulge\_JOHNSON\_B*        & 90& burst\_JOHNSON\_B*        \\
 29& disc\_JOHNSON\_V*        & 60& bulge\_JOHNSON\_V*        & 91& burst\_JOHNSON\_V*        \\
 30& disc\_JOHNSON\_H*        & 61& bulge\_JOHNSON\_H*        & 92& burst\_JOHNSON\_H*        \\
 31& disc\_JOHNSON\_I*        & 62& bulge\_JOHNSON\_I*        & 93& burst\_JOHNSON\_I*        \\
 32& disc\_JOHNSON\_J*        & 63& bulge\_JOHNSON\_J*        & 94& burst\_JOHNSON\_J*        \\
 33& disc\_JOHNSON\_K         & 64& bulge\_JOHNSON\_K         & 95& burst\_JOHNSON\_K         \\
 34& disc\_JOHNSON\_R*        & 65& bulge\_JOHNSON\_R*        & 96& burst\_JOHNSON\_R*        \\
 35& disc\_UV\_FOCA\_highres* & 66& bulge\_UV\_FOCA\_highres* & 97& burst\_UV\_FOCA\_highres* \\
 36& disc\_IRAS\_12mic*       & 67& bulge\_IRAS\_12mic*       & 98& burst\_IRAS\_12mic*       \\
 37& disc\_IRAS\_25mic*       & 68& bulge\_IRAS\_25mic*       & 99& burst\_IRAS\_25mic*       \\
 38& disc\_IRAS\_60mic*       & 69& bulge\_IRAS\_60mic*       &100& burst\_IRAS\_60mic*       \\
 39& disc\_IRAS\_100mic*      & 70& bulge\_IRAS\_100mic*      &101& burst\_IRAS\_100mic*      \\
 40& disc\_ISOCAM\_15mic*     & 71& bulge\_ISOCAM\_15mic*     &102& burst\_ISOCAM\_15mic*     \\
 41& disc\_ISOPHOT\_170mic*   & 72& bulge\_ISOPHOT\_170mic*   &103& burst\_ISOPHOT\_170mic*   \\
 42& disc\_MIPS\_024mic*      & 73& bulge\_MIPS\_024mic*      &104& burst\_MIPS\_024mic*      \\
 43& disc\_MIPS\_070mic*      & 74& bulge\_MIPS\_070mic*      &105& burst\_MIPS\_070mic*      \\
 44& disc\_SPIRE\_250mic*     & 75& bulge\_SPIRE\_250mic*     &106& burst\_SPIRE\_250mic*     \\
 45& disc\_SPIRE\_350mic*     & 76& bulge\_SPIRE\_350mic*     &107& burst\_SPIRE\_350mic*     \\
 46& disc\_SPIRE\_500mic*     & 77& bulge\_SPIRE\_500mic*     &108& burst\_SPIRE\_500mic*     \\
 47& disc\_UV\_1600Ang*       & 78& bulge\_UV\_1600Ang*       &109& burst\_UV\_1600Ang*       \\
 48& disc\_UV\_1500Ang*       & 79& bulge\_UV\_1500Ang*       &110& burst\_UV\_1500Ang*       \\
 49& disc\_PLANCK\_550mic*    & 80& bulge\_PLANCK\_550mic*    &111& burst\_PLANCK\_550mic*    \\
 50& disc\_PLANCK\_850mic*    & 81& bulge\_PLANCK\_850mic*    &112& burst\_PLANCK\_850mic*    \\
 51& disc\_PLANCK\_1380mic*   & 82& bulge\_PLANCK\_1380mic*   &113& burst\_PLANCK\_1380mic*   \\
 52& disc\_PLANCK\_2100mic*   & 83& bulge\_PLANCK\_2100mic*   &114& burst\_PLANCK\_2100mic*   \\
 53& disc\_PLANCK\_3000mic*   & 84& bulge\_PLANCK\_3000mic*   &115& burst\_PLANCK\_3000mic*   \\
 54& disc\_IRAC\_3\_6mic*     & 85& bulge\_IRAC\_3\_6mic*     &116& burst\_IRAC\_3\_6mic*     \\
 55& disc\_IRAC\_4\_5mic*     & 86& bulge\_IRAC\_4\_5mic*     &117& burst\_IRAC\_4\_5mic*     \\
 56& disc\_IRAC\_5\_8mic*     & 87& bulge\_IRAC\_5\_8mic*     &118& burst\_IRAC\_5\_8mic*     \\
 57& disc\_IRAC\_8\_0mic*     & 88& bulge\_IRAC\_8\_0mic*     &119& burst\_IRAC\_8\_0mic*     \\
		\end{tabular}
	\end{center}
	\caption{List of characters. \textit{mstar}, \textit{mcold} and \textit{mcoldz} stand respectively for the masses of stars, gas and metals. \textit{rgal} is the component radius, \textit{speed} its rotation speed, \textit{tdyn} the dynamical time, and \textit{inst\_sfr} the instantaneous star formation rates. Magnitudes (characters 27 to 119, starred) are relative to the K band of each component. See text for more details.}
	\label{tabchar}
\end{table}

Each galaxy is identified by a specific number at each timestep of the simulation. Each galaxy is the product of one or more galaxies of the previous step and one or more evolutionary processes that occurred since the previous step. The entire genealogy of each galaxy is thus known. The sample was selected among galaxies at the present epoch (redshift=0) of the simulation. For any given galaxy, we counted the total number of merging events that lead to its formation since the birth of the very first of its ancestors. We arbitrarily selected a number of galaxies with 1, 2, 3, 4, 5, 6, 8 and 10 mergers (5 galaxies each) and with 15, 20 and 22 mergers (1 galaxy each) for a total of 43 galaxies. We also noted the number of major mergers, implying two galaxies with a mass ratio larger than 1/3. In order to visualize quickly some correlations on the final tree, we named the galaxies as XXcYAAN where XX is the total number of mergers, Y the number of major events and AA stands for central (CE), satellite (SA) or field (CH) galaxies. N merely identifies galaxies having the same XX and AA.

They are described by 119 parameters which we assumed to be evolutive characters (Table~\ref{tabchar}). There are the total bolometric luminosity and the total bolometric IR (infrared) flux. Then, for each of the three components of galaxies (disc, bulge and burst), there are some size, mass and kinematic data, but principally photometric values from the ultraviolet to the far infrared. Like in \cite{paperI}, all magnitudes are relative to the K band, this last value giving the relative heights of the spectra or relative brightnesses between galaxies. We replaced the size of the bulge (bulge\_rgal) with the ratio between sizes of bulge and disc (bulge\_rgal/disc\_rgal). The dynamical time $tdyn$ is the time taken for material at the half-mass radius to reach the opposite side of the galaxy (disk component) or its centre (bulge and burst components), whereas the instantaneous star formation rate is derived from this dynamical time, the mass of the cold gas and a prescribed star formation efficiency, all taken at the last time substep of the simulation (Hatton et al. \cite{galics1}).

Total magnitudes for the galaxies, merely computed by adding intensities of the three components, were not included in the analysis. They are certainly less precise to describe evolution of galaxies and might introduce some redundancy, that is some artificial overweighting of some characters. However, they can be projected onto the final cladograms for interpretation purposes (see Sect.~\ref{discussion}).

\section{Astrocladistics analysis}
\label{cladanal}

For a detailed presentation of the method, the reader should refer to \cite{paperI} and references therein. Only ingredients particular to the analysis performed in this second paper are described in this Section.

\subsection{The matrix}

Following results from \cite{paperI}, spectral characters were included in the matrix as colour magnitudes with respect to the K~band for each component. This band provides relative brightness between galaxies, while colours indicate the evolutionary state for physical and chemical constituents. The original table is available on request.

\subsection{Coding of the matrix}

The evolution of characters is described by 8 discrete states which are regular bins in colour magnitudes. In case one magnitude for a given character is undetermined (value ``99'' in the initial matrix standing for null flux), code 7 was attributed and the other entries coded from 0 to 6. The coded matrix is presented in Table~2.

\subsection{Additional constraints}

An ordered (Wagner) evolution was imposed on the characters, so that they are supposed to evolve smoothly with time: changes between two adjacent states are more parsimonious than between distant ones. This assumption is physically plausible and we found that it significantly improves the robustness of the final tree.

\subsection{Outgroup}
\label{outgroup}

The choice of an outgroup is required to root the cladogram or to orientate the arrow of time. However, this is always a difficult question, particularly when the phylogenesis of the group under study is not understood. This is obviously the case for galaxies because astrocladistic galaxy classification is still in its infancy. 

Nonetheless, the absence of any identified outgroup does not prevent a phylogenetic analysis to be made because it still provides invaluable results on relationships and groupings of objects from an evolution point of view. In this paper, we will try to consider 01c0CH2, 01c0CH3, 01c0CH4 as a priori possible outgroups (less diversified objects) for the group under study since they result from only one non-major merger. Indeed, this argument looks like an a priori hypothesis stating that merger events are a dominant driver of diversification. In view of Sect.~\ref{biology}, this is still to be demonstrated, and the present analysis does not support this point (Sect.~\ref{results}). Our choice is thus dictated by display purposes only.

\subsection{Heuristic quest for the best trees}

We used the PAUP4b10* package (Swofford~\cite{paup}) on Linux PC computers to perform all calculations shown in this paper. The maximum parsimony criterion was chosen. Since we do not know anything about evolutionary relationships of the objects of our sample, bootstrap values (\cite{paperI}) were optimized by removing object after object until we reach a satisfactory tree, i.e. a tree displaying highly supported nodes. 

\subsection{Assessment of the phylogenetic signal in the data}

The robustness of the final result was decided from bootstrap values and confirmed by decay Bremer indices (see \cite{paperI} for details).


\section{Results}
\label{results}

\subsection{The best cladogram}

As already mentioned in \cite{paperI}, burst characters are probably not very pertinent for evolution since they concern a temporary component of galaxies. Comparison runs showed us that results without them are more robust. The high variability of burst characters is confirmed on their projections on the final cladogram (not shown in this paper).
The analysis presented afterwards is thus performed with the 80 remaining characters.

   \begin{figure}
   \centering
   \includegraphics[width=17 true cm]{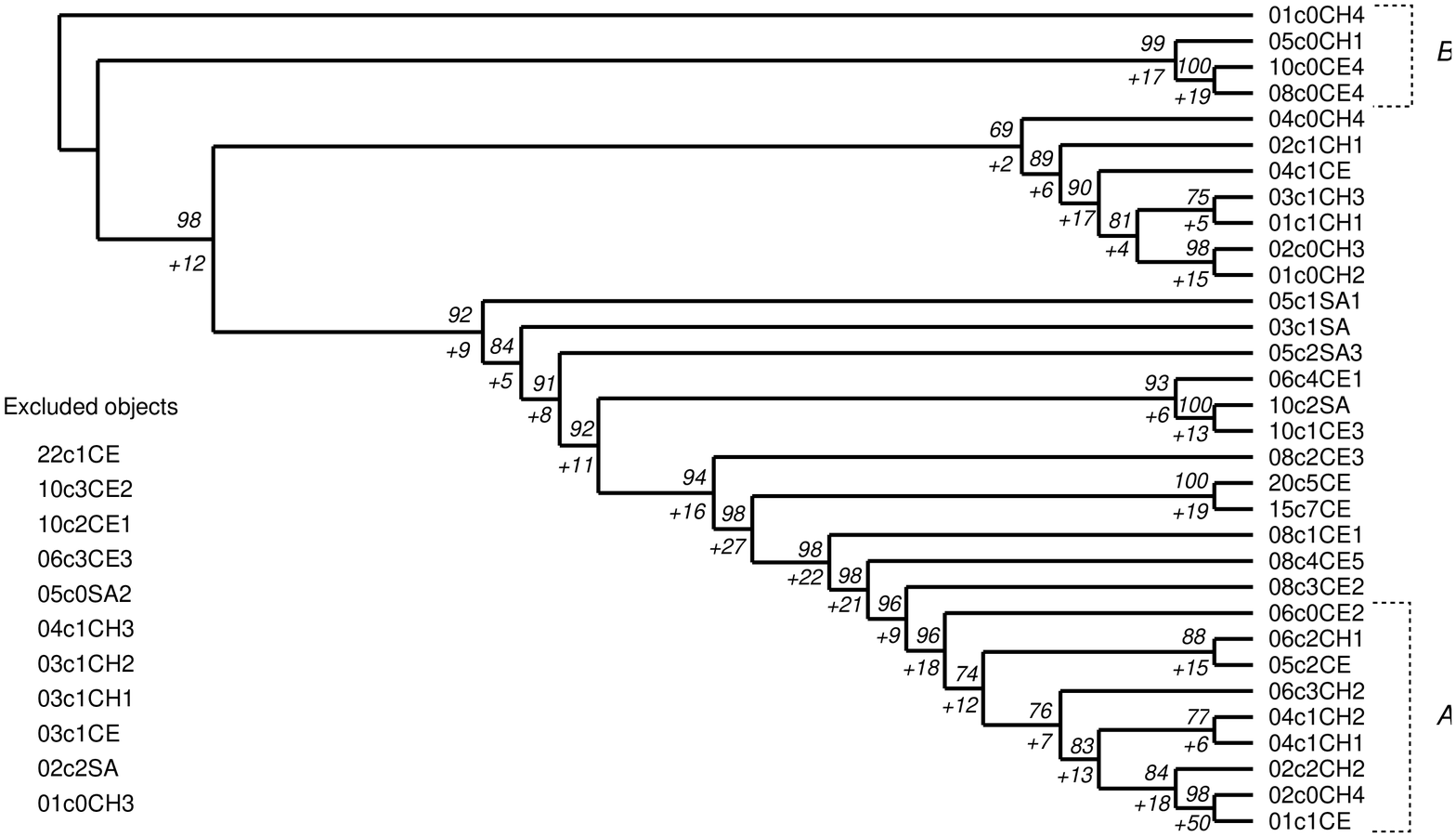}
      \caption{Best cladogram obtained without burst characters and after step by step removal of the indicated galaxies. See text for galaxy nomenclature. Letters A and B correspond to groups mentioned in the text. Numbers to the left of each node are bootstrap (above) and decay (below with a plus sign) values. The number of steps for this tree is 1364. Consistency Index (CI) = 0.41, Retention Index (RI) = 0.71 (see \cite{paperI} for details).
              }
         \label{cladogram}
   \end{figure}
    
The best tree was found after excluding 11 objects. They must be considered as objects perturbing the cladogram probably because they do not share common evolution patterns with the other ones. We stopped the optimization process when we reached a point where excluding one or two more objects or some more characters was not improving the bootstrap value for any node. We consider the result as highly significant. The cladogram with 32 objects and 80 ordered characters is shown in Fig.~\ref{cladogram}. The 11 excluded objects are not found to build a group (no robust tree). 

The outgroup (see Sect.~\ref{outgroup}) was chosen to be 01c0CH4 for a better visualization of the cladogram than with 01c0CH2, whereas 01c0CH3 belongs to the excluded objects. This already shows that basing evolution (diversification) only on merger events is not a valid approach.

The analysis of the entire initial matrix revealed a partly resolved tree but this was poorly supported. 
Even if the result shown is obtained for 32 objects only, it is still remarkable to find such a cladogram, well supported by excellent bootstrap and decay values, for an arbitrarily chosen sample of galaxies (see a discussion in \cite{paperI}). This really tells us that branching evolution is the dominant diversification mechanism, and that mergers do not destroy the tree-like organization, at least in the GALICS simulations.

There might be a trend toward a correlation between diversification and number of mergers. Among the 9 galaxies with no major merger (labelled *c0*), 7 are grouped in the upper part of the cladogram, 06c0CE2 and more noticeably 02c0CH4 being at the opposite end (the most diversified objects from 01c0CH4). In addition, three galaxies (05c0CH1, 08c0CE4, 10c0CE4) in the upper part have a number of mergers significantly higher than the other ones. Except for these three objects, there is a regular increase of the total number of mergers downward on the cladogram, but after 20c5CE a regular decrease occurs. Objects with more major mergers are also grouped together, roughly in the middle of the cladogram. But there is no regular increase with evolution. We conclude that, even if mergers are a driver of diversification, they are not the principal one and cannot be considered as a reliable evolution clock. 

The two galaxies with the highest number of mergers and major mergers (15c7CE and 20c5CE) are clearly grouped together (bootstrap of 100), obviously sharing similar histories as compared to all the other ones. There seems to be a loose trend for the number of major mergers to increase with the total number of mergers, the corresponding galaxies becoming more central objects. There is a noticeable exception (22c1CE) with only 1 major event for 22 mergers in total and which belongs to the excluded objects.

\subsection{Evolution of characters}
\label{charevol}
   \begin{figure}
   \centering
   \includegraphics[width=16 true cm]{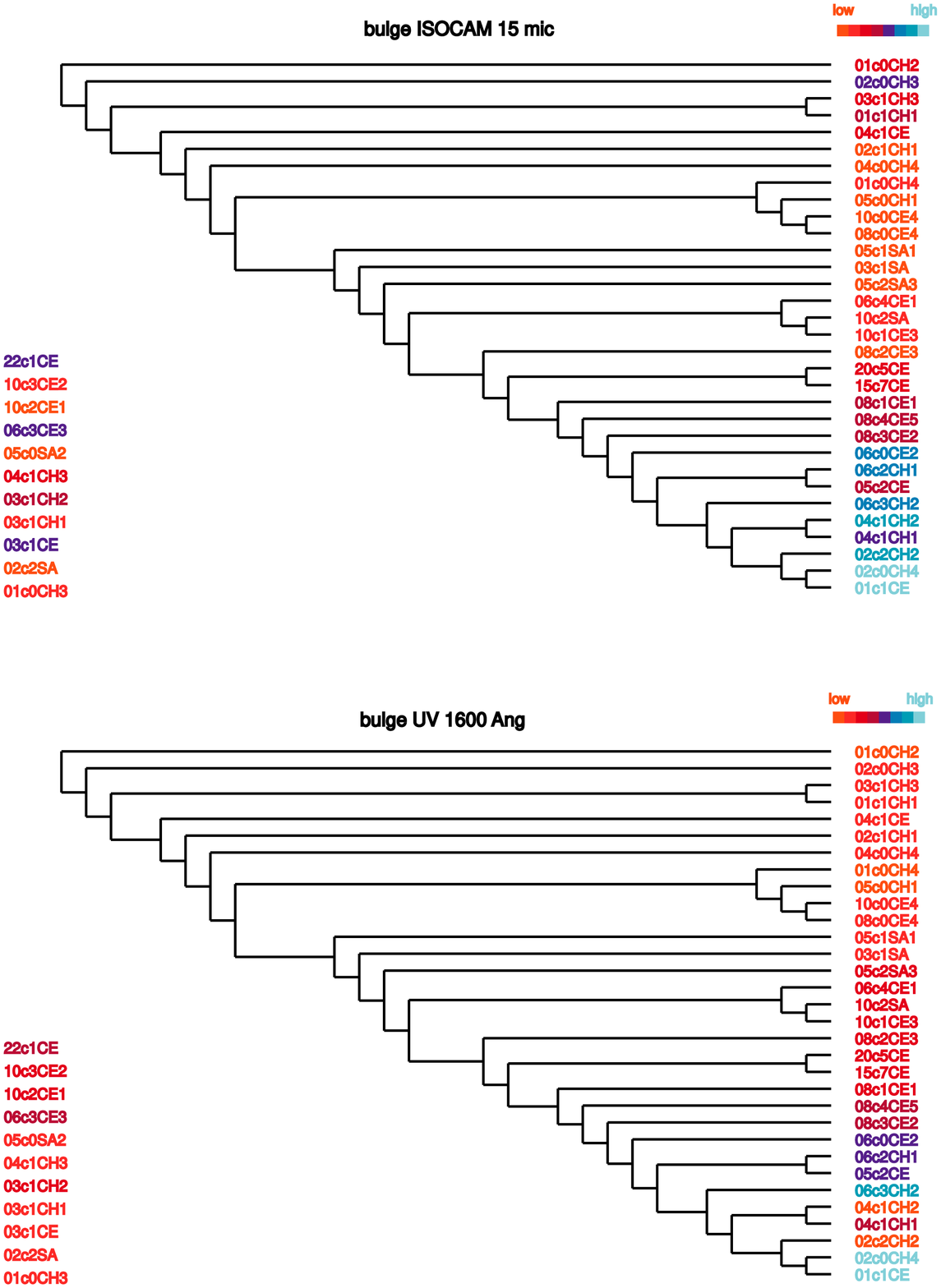}
      \caption{Cladogram of Fig.~\ref{cladogram} with projected characters in colour codes indicated on top right. 'Low' and 'high' refer to magnitudes with respect to the $K$ band (see Table~\ref{tabchar}).
              }
         \label{bulge}
   \end{figure}
   
   \begin{figure}
   \centering
   \includegraphics[width=16 true cm]{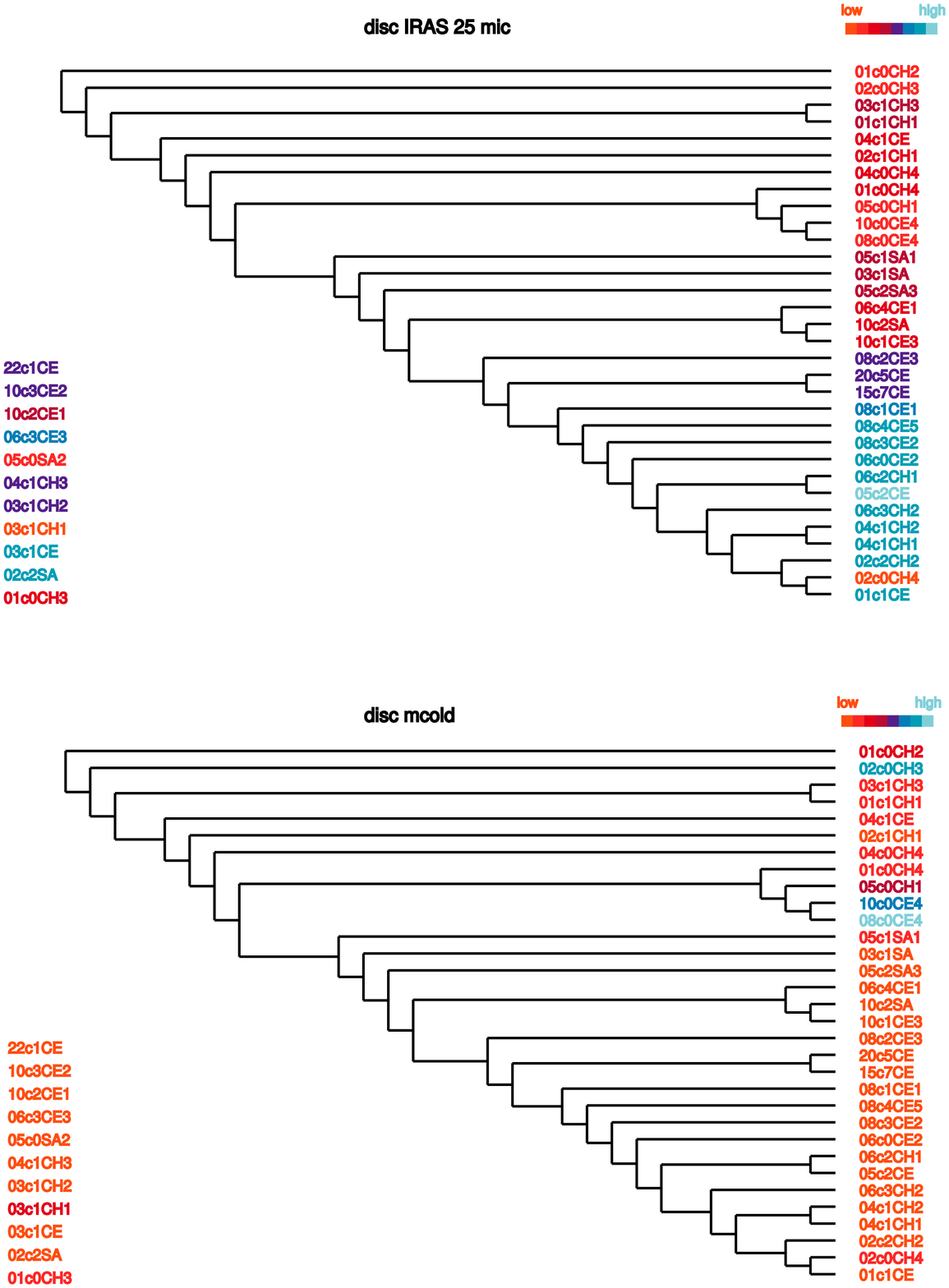}
      \caption{Same as Fig.~\ref{bulge} for two disc characters.
              }
         \label{disc}
   \end{figure}

In the previous section, it was noticed that organization on the cladogram is not entirely due to merger processes. So what causes the hierarchy? We present in Fig.~\ref{bulge} and  Fig.~\ref{disc} projections of four characters in colour codes (with respect to the $K$ band, see Table~\ref{tabchar}). They help understand unique properties of groups of galaxies. It is found that there is a group of galaxies at the bottom (identified as \textit{A} in Fig.~\ref{cladogram}) characterized by very high values of near infrared and UV (ultraviolet) colours (with respect to the K band) for the bulge component. In addition, their disc has also a very high colour value in the near infrared. Finally, the cold gas mass of the disc is decreasing regularly from the top toward the bottom of the cladogram as shown, with a group (\textit{B} on Fig.~\ref{cladogram}) having high masses.

These properties alone explain the organization found for this sample. The other characters generally behave somewhat more erratically in this picture. With this information, it is possible to understand the events that create such character evolutions, hence galaxy diversity, by relating the observable to the basic constituents (stars, gas and dust). This is beyond the scope of this work and is reserved for further and more complete studies of GALICS galaxies, as well of course of real ones.

\section{Discussion}
\label{discussion}

The cladogram presented in the previous section is a synthetic way to visualize a set of data and hypotheses. Results should be discussed in view of these inputs. Concerning the choice of the objects, as already noticed, it is remarkable to find a well resolved tree with 32 galaxies while the total sample (43) has been arbitrarily selected among thousands of possible candidates. This means that the majority of our sample object classes have a common ancestor (ancestral species). Their very first progenitors, born at different times and places in the Universe, were made up with material of comparable histories and underwent similar physical processes. Thus the Universe was not very inhomogeneous, but one should remember that this is a simulated Universe.

To obtain a very robust tree, 11 objects have been excluded. Several explanations can be invoked. These galaxies can be too distant in the evolution (too much diversified) for any valuable relationships with the other to be discovered. More characters could be needed to establish the global phylogeny. It is possible also that there are some true hybrid specimens that cannot easily fit in a cladogram with the others. Clearly, more galaxies should be studied and further work is necessary before understanding these objects in view of phylogenetic classes still to be defined.

Individual galaxies are representative of classes. This is a one-galaxy one-class assumption which should be checked: if two galaxies are very similar in their coded characters, they could define a class. Quantitative values have been divided into 8 bins to build the coded matrix. Nevertheless, it is possible to reduce this number, in order to avoid over-resolution and allow for some initial groupings. In our case, this gave worse results, without revealing galaxies with similar sets of coded characters. As presented in \cite{paperI}, comparing two objects in evolution can be done in several ways. Cladistic analysis is one way, but it defines classes a posteriori, using cladograms, and not a priori. This is a step by step process and a long term goal of astrocladistics. 

For better objectivity of the methodology of classification (see \cite{paperI}), all available observables should be a priori included. The cladistic analysis reveals their pertinence and their behaviour regarding evolution. The fundamental point is that the decision of excluding some characters is to be made \emph{afterwards}, based on objective and transparent arguments provided by the analysis itself, associated statistical tools and interpretation of the cladogram. In our case, including burst characters clearly hampered the convergence toward a robust tree. Projections on the cladograms unambiguously show their high variability with time (figures available on request). 

The hierarchical organization is mainly explained by infrared properties (Sect.~\ref{charevol}). This has naturally a strong impact on our understanding of the physics of the galaxy evolution and diversity (since infrared radiation mainly indicates temperature of the dust). This could also be partly caused by redundancy of this information in the matrix. But how can we be sure not to artificially weight characters and thus influence important evolution indicators? A way around this difficulty would be to compare cladograms obtained with different weight hypotheses, and check the overall astrophysical consistency of the resulting interpretations. We reserve such an analysis for further studies with larger samples within GALICS. It is important to realize that a cladogram is never definitive. For instance, future discoveries will bring new information in the initial matrix, and the cladogram will change accordingly (see \cite{paperI}).

The GALICS simulations do not take interactions and ejection-sweeping into account. Our results show that diversification occurs in a hierarchical way and mergers obviously play a role. Because they do not seem to constitute an evolutionary clock, they are not the dominant process for galaxy evolution so that mergers, accretion, secular evolution and also assembling compete for diversification.

The astrophysical goals of this paper were twofold: the formalization of formation and diversification processes, and the role of mergers in branching evolution. The first point can now be employed in astrocladistic analyses of real galaxies. The second can only be addressed by simulated samples as we did. Future simulations, including interactions between galaxies and kinematic information, will be invaluable for making precise estimations of the respective importance of the five formation processes identified in this paper, by comparing phylogenetic analyses of both simulated and real samples.

Can we apply astrocladistics to real galaxies? Our works in \cite{paperI} and here demonstrate that there are absolutely no obstacles: characters used are real observables. Indeed, this has already been undertaken for samples of Dwarf Galaxies of the Local Group and of Virgo galaxies (see Fraix-Burnet ~\cite{fraix} for an overview). The more characters we have, the more robust and the larger the galaxy tree is. This series of papers on simulated galaxies show that photometric characters are well suited. For real galaxies, other parameters are available as well, particularly on chemical composition and kinematics. This last category of observables is particularly important because it is a potential tracer of past interactions and mergers. In addition, the apparent morphology of galaxies, a subjective and qualitative character used in the Hubble classification, is essentially caused by and included in kinematic properties, which are objective and quantitative descriptors.


\section{Conclusions}
\label{conclusions}

Astrocladistics relies on two basic requirements presented in \cite{paperI} and this paper: galaxies and their evolution are defined and described by their basic constituents (stars, gas and dust), diversification is due to branching evolution. The success of an analysis is judged from objective statistical methods and from the astrophysical interpretation of the cladogram. 

The analysis of a sample of simulated galaxies shows that mergers do not destroy the hierarchical organization of diversity. This is because it is not such a different event from accretion, except for its violence. The basic constituents are mixed together and not replaced. Mergers thus cannot be paralleled by hybridization leading to reticulated evolution in biology. In conclusion, the five galaxy formation processes identified in this paper participate in a branching evolution. 

It is difficult to draw general conclusions on the phylogeny of galaxies from the analysis presented in this paper and it was not its objective. Interactions between galaxies are not taken into account in the GALICS simulation. This formation process is certainly the most frequent one, and being in general very violent, it is expected to be one of the main drivers of diversification.

A lot more studies are possible with the GALICS database, particularly to help understand how diversity occurs in such simulations. It is also an invaluable tool to learn more on astrocladistic methodology and interpretation of its results. Finally, it helps in handling samples of real galaxies for which we have no historical information and possibly less available characters. In both papers of this series, the results unambiguously validate the approach. Ongoing positive results on Dwarfs Galaxies of the Local Group and Virgo galaxies (Fraix-Burnet~\cite{fraix}) show that this conclusion is not limited to simulated galaxies. The final game is here: beyond a necessary new taxonomic classification, is the establishment of the tree of galaxies. Many more astrocladistic studies are required, as well as gathering as many observational galaxy descriptors as possible. 

Current and future big surveys (like Sloan Digital Sky Survey, Vimos VLT Deep Survey, Great Observatories Origins Deep Survey, ...) collecting photometric and spectroscopic data on huge samples of galaxies at all redshifts seem particularly well suited for astrocladistics. Such large data sources will be integrated within the international Virtual Observatory project (http://www.ivoa.net/) that will provide large numbers of characters for large samples of diversified galaxies.

It is also true that, even if, somewhat like in biology, the environment plays a crucial role in galaxy evolution (accretions, interactions and mergers), galaxies set new conditions for cladists: continuous data, comprising error bars and upper/lower limits, violent generation of species, huge amount of objects for not that many characters, instantaneous galaxy diversity that might decrease because of gravity leading to fewer and fewer objects. In addition, the detailed evolutionary physical and chemical processes are basically understood, and objects of the past (high redshift galaxies) will be observed with future giant telescopes at the same detail as present day galaxies.

\begin{acknowledgements}
The authors wish to thank the two referees for very interesting and helpful comments on the manuscript.
      This work uses the GalICS/MoMaF Database of Galaxies (http://galics.iap.fr). We warmly thank Bruno Guiderdoni and Jeremy Blaizot from the GALICS project for invaluable help and discussions. All the GALICS team should be thanked for providing such a useful database. Emmanuel Davoust kindly read the manuscript and helped us much in improving it. Most computations of the results presented in this paper were performed under the CIMENT project in Grenoble. The contribution of EJPD is publication 2006-002 of the Institut des Sciences de l'\'Evolution de Montpellier (UMR 5554 - CNRS).
\end{acknowledgements}

   \begin{figure}
   \centering
   \includegraphics[width=16 true cm]{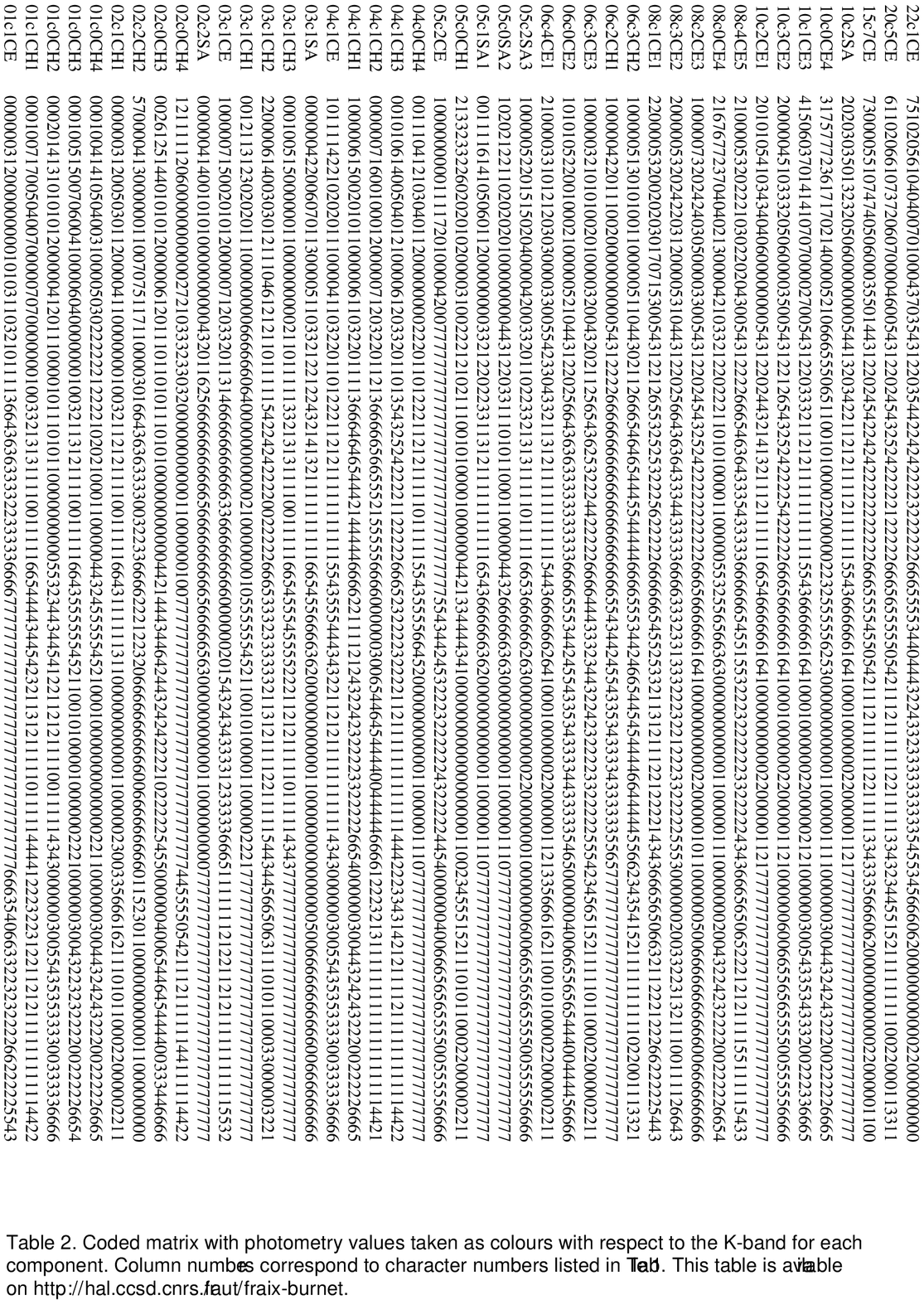}
   \end{figure}

\end{document}